# Mechanism of the single-pulse ablative generation of laser induced periodic surface structures

Maxim V. Shugaev,[1] Iaroslav Gnilitskyi,[2] Nadezhda M. Bulgakova,[3,4] and Leonid V. Zhigilei [1]

[1] University of Virginia, Department of Materials Science and Engineering, 395 McCormick Road, Charlottesville, VA 22904-4745, USA

[2] DISMI, University of Modena and Reggio Emilia (UNIMORE), 2 via Amendola, Reggio Emilia 41122, Italy

[3] HiLASE Centre, Institute of Physics ASCR, Za Radnicí 828, 25241 Dolní Břežany, Czech Republic

[4] S.S. Kutateladze Institute of Thermophysics SB RAS, 1 Lavrentyev Ave., Novosibirsk 630090, Russia

**Abstract**

One of the remarkable capabilities of ultrashort polarized laser pulses is the generation of laser-induced periodic surface structures (LIPSS). The origin of this phenomenon is largely attributed to the interference of the incident laser wave and surface electromagnetic wave that creates a periodic absorption pattern. Although, commonly, LIPSS are produced by repetitive irradiation of the same area by multiple laser pulses in the regime of surface melting and resolidification, recent reports demonstrate the formation of LIPSS in the single pulse irradiation regime at laser fluences well above the ablation threshold. In this paper, we report results of a large-scale molecular dynamics simulation aimed at providing insights into the mechanisms of single pulse ablative LIPSS formation. The simulation performed for a Cr target reveals an interplay of material removal and redistribution in the course of spatially modulated ablation, leading to the transient formation of an elongated liquid wall extending up to ~600 nm above the surface of the target at the locations of the minima of the laser energy deposition. The upper part of the liquid wall disintegrates into droplets while the base of the wall solidifies on the timescale of ~2 ns, producing a ~100 nm-long frozen surface feature extending above the level of the initial surface of the target. The properties of the surface region of the target are modified by the presence of high densities of dislocations and vacancies generated due to the rapid and highly nonequilibrium nature of the melting and resolidification processes. The insights into the LIPSS formation mechanisms may help in designing approaches for increasing the processing speed and improving the quality of the laser-patterned periodic surface structures.

## 1. Introduction

One of the remarkable capabilities of ultrashort (femto- and picosecond) polarized laser pulses is imprinting of light polarization direction on a material surface in the form of laser-induced periodic surface structures known as LIPSS (see [1–3] and references therein). These structures are already used and have a potential to be applied in various fields, including control over surface wettability [4,5], enhancing light absorption [5], improving adhesive [6] and tribological [7] properties of surfaces, surface marking [8], controlled growth and migration of cells [9], etc. The origin of the universal phenomenon of LIPSS formation, observed on metals, semiconductors, dielectrics, and polymers irradiated at a broad range of laser wavelengths and pulse durations, is largely attributed to the interference of the incident laser wave and a surface scattered electromagnetic wave, thus creating a periodic pattern of the absorbed laser energy along the irradiated surface, as established by Sipe *et al.* for metals [10] and generalized to other materials by Bonse *et al.* [11]. A periodic (modulated) absorption of laser energy by electrons during the laser pulse is only the first stage of the LIPSS formation. This stage results in a modulation of the lattice temperature along the surface, which has been shown to preserve up to tens of picoseconds even on surfaces of metals with high thermal conductivity [12]. While such timescale is sufficient for triggering the material melting and ablation [13], the longer-term post-irradiation processes leading to the material redistribution and/or removal within/from the surface layer of the target must play a paramount role in the final imprinting of the LIPSS on the surface.

Usually LIPSS imprinting on material surfaces is performed by laser pulses with relatively low laser fluence, near or only slightly above the ablation threshold [1,2,8,11,14]. This requires dozens or even hundreds of laser pulses coupled to the same irradiation spot on the surface to produce the periodic surface relief. It was believed that an increase of the laser fluence well above the ablation threshold would result in disordering or even complete erasure of the periodic structures by the recoil pressure of the ablation plume. It was recently shown, however, that a well-pronounced LIPSS pattern can be imprinted on the surface with a single or very few laser pulses coupled to the same irradiated area at fluences well above the ablation threshold [15–17]. In particular, in Ref. [17] the high-quality LIPSS were written on a silicon surface by fast laser scanning with overlap of only 2-3 pulses per irradiation spot. The mechanisms of LIPSS formation in the regimes of developed ablation are not established yet. It was speculated [17] that periodic ablation, which replicates the laser energy absorption pattern, and material relocation due to recoil pressure of the ablation products could cause the periodic surface rippling. In this paper, we report the results of a large-scale molecular dynamics simulation that suggest an alternative mechanism of the single pulse ablative LIPSS formation, where the material redistribution in the lower part of the ablation plume plays an essential role in the formation of frozen surface protrusion in the regions of reduced laser energy deposition. Detailed structural analysis of the frozen protrusions reveals a high concentration of crystal defects that are likely to have important implications on the mechanical and chemical properties of the nanostructured surfaces.

## 2. Computational setup

The simulation reported in this paper is performed with a hybrid computational model combining continuum level description of laser excitation of conduction band electrons and following electron-phonon equilibration based on Two Temperature Model (TTM) [18] and classical molecular dynamics (MD) representation of the non-equilibrium dynamics of laser-induced phase transformations and material decomposition. A complete description of the computational model is provided elsewhere [19,20], and below we only delineate parameters of the computational setup specific for the simulation reported in this paper.

A schematic representation of the computational system is illustrated in Figure 1. The initial dimensions of the TTM-MD domain are 260 nm × 43 nm × 87 nm, which corresponds to 81 million atoms. The interatomic interactions are described by the embedded atom method (EAM) potential parametrized for Cr [21]. The potential provides a relatively accurate and computationally inexpensive description of Cr experimental properties, including lattice parameter, cohesive energy, elastic constants and their temperature dependence, melting temperature, and vacancy formation energy. The electronic heat transfer in the deeper part of the target, where no structural changes take place in response to the laser irradiation, is described by the conventional TTM. The depth covered by the TTM is chosen to be 1 μm to ensure a negligible temperature change at the bottom of the computational system by the end of the simulation. Periodic boundary conditions are applied along $x$ and $y$ directions (parallel to the irradiated (001) surface of the target), a free boundary is used at the top surface of the target, and a new Langevin Non-Reflecting Boundary (LNRB) condition, described in Appendix A, is imposed at the bottom of the TTM-MD part. The LNRB condition ensures nonreflecting propagation of laser-induced non-planar pressure wave into the bulk of the target, mimicking the elastic response of an infinitely thick target. This approach is suitable for simulation of experimental conditions where the reflection of the pressure waves from the back surface of the irradiated target does not have any significant effect on processes occurring in the vicinity of the irradiated surface. After 250 ps, the dynamic relaxation of laser-induced stresses in the TTM-MD part of the system is completed and the LNRB is replaced with a rigid boundary condition. Before laser irradiation, the system is thermalized at 300 K for 50 ps.

The thermophysical properties of Cr entering the TTM equations are as follows. The electron heat capacity of Cr is approximated as $C_e = \gamma T_e$ with $\gamma = 194$ Jm$^{-3}$K$^{-2}$ [22,23]. Constant values of the electron-phonon coupling factor, $G = 4.2 \times 10^{17}$ Wm$^{-3}$K$^{-1}$ [24], and the lattice heat capacity in the TTM part of the model, $C_l = 3.23 \times 10^6$ Jm$^{-3}$K$^{-1}$ [25], are assumed in the calculations. The temperature dependence of the electron thermal conductivity is approximated by the Drude model relationship, $K_e(T_e,T_l) = v^2 C_e(T_e) \tau_e(T_e,T_l)/3$, where $C_e(T_e)$ is the electron heat capacity, $v^2$ is the mean square velocity of the electrons contributing to the electron heat conductivity, approximated in this work as the Fermi velocity squared, $v_F^2$, and $\tau_e(T_e,T_l)$ is the total electron scattering time. The latter is defined by the electron-electron scattering rate, $1/\tau_{e-e} = AT_e^2$, and the electron-phonon scattering rate, $1/\tau_{e-ph} = BT_l$, so that $1/\tau_e = AT_e^2 + BT_l$. The value of the coefficient $A = 2.66 \times 10^6$ K$^{-2}$s$^{-1}$ is estimated within the free electron model [26]. Similar to Ref. [27], The coefficient $B$ is described as a function of the lattice temperature and the phase state of the material, so that the experimental temperature dependences of thermal conductivity of Cr under conditions of electron-phonon equilibrium [28] are reproduced for both solid and liquid states. In particular, this description accounts for the two-fold drop of thermal conductivity as the temperature increases from 300 K to the melting point.

The absorption of laser energy is represented through a source term added to the TTM equation for the electron temperature [19]. The source term has a temporal Gaussian profile corresponding to 200 fs pulse and accounts for the exponential attenuation of the deposited laser energy with depth. The optical absorption depth of Cr is equal to 9.6 nm at the wavelength 258 nm [29]. Similar to Ref. [12], the absorbed laser fluence is spatially modulated along the $x$ direction to represent the periodic absorption pattern generated by the interference of the incident laser wave and surface electromagnetic wave. The period of the sinusoidal modulation is taken to be 260 nm, which matches the size of the computational domain in $x$ direction. The maximum and minimum of absorbed fluences (at the boundaries and in the middle of the computational domain shown in Figure 1, respectively) are 2000 to 3000 J/m$^2$, which corresponds to 20% modulation of the laser energy deposition with respect to the average level of 2500 J/m$^2$. Such level of modulation of the absorbed laser energy is predicted to result from periodic generation of free carriers [30] in silicon, and is also expected for metals exhibiting a pronounced decrease of reflectivity with heating of the electron subsystem [31].

## 3. Results and discussion

The mechanisms of the single pulse LIPSS formation in the regime of laser ablation is investigated in a large-scale TTM-MD simulation performed for a Cr target irradiated by a 200 fs laser pulse. The irradiation is assumed to produce a spatially modulated energy deposition with the average absorbed fluence of 2500 J/m$^2$, which is approximately 70% above the phase explosion threshold for Cr target.

The initial response of the Cr target to the modulated laser energy deposition is illustrated in Figure 2, where the evolution of density distribution and material flow are shown for the initial 100 ps after the laser pulse. Laser excitation and rapid electron-phonon equilibration leads to the heating of the top region of the target and induces an upward material expansion as early as 5 ps. Since the absorbed laser fluence is the lowest in the middle of the computational system and highest on the sides, the temperature and pressure gradients have lateral components and drive material rearrangement towards the center, as demonstrated by the deviation of the arrows showing the material outflow from the vertical direction. Meanwhile, at the bottom part of the system, smaller downward-pointing arrows are also inclined toward the center, signifying the nonplanar nature of the pressure wave propagating into the bulk of the target and highlighting the need for the modified nonreflecting boundary condition applied at the bottom of the TTM-MD domain (see Appendix A). The following density snapshots in Figure 2 illustrate the rapid expansion and following decomposition of the material overheated above the limit of its thermodynamic stability. The higher vapor pressure generated at the periphery of the computational cell (the area of the higher energy deposition) drives the vapor and liquid droplets to the central part of the system, where a high-density region evolving into a liquid wall is generated.

The formation of the liquid wall can be clearly seen in the snapshots of atomic configurations shown in Figure 3 for a longer time of up to 2100 ps. The base of the liquid wall emerges from the foamy structure of interconnected liquid regions generated in the lower part of the ablation plume, while the upper part forms through the coalescence of small liquid droplets pushed toward the center by the lateral component of the vapor pressure gradient. Moving by inertia upward, the liquid wall elongates and becomes thinner. The wall reaches its maximum length of ~600 nm by ~1300 ps, when it starts to decompose into separate liquid droplets. Meanwhile, the rapid electron heat conduction to the bulk of the target undercools the molten surface region and causes the upward motion of the solid-liquid interface. The solidification front reaches the base of the liquid wall by ~500 ps, and the surface region of the target completely solidifies at 2100 ps, shortly after the last liquid droplets are separated from the wall. As can be seen from Figure 3, the solidification of the lower part of the transient liquid wall leads to the formation of an elongated protrusion with the height of approximately 100 nm. Note that the amount of material removed from the trough area of the modified surface corresponds to the ablation depth of ~40 nm, while the frozen protrusion extends above the level of the original surface of the target. The latter computational prediction is consistent with experimental observations of Refs. [16,31], where the LIPSS pattern generated by single femtosecond pulse irradiation of Au has a characteristic shape of frozen wall-like structures similar to the ones in Figure 3. The simulation results demonstrate that the height and shape of the protrusion is defined by the competition between the disintegration of the liquid wall into droplets and the propagation of the solidification front. The latter accelerates with increasing undercooling of the crystal-liquid interface, reaches its maximum velocity of ~65 m/s when the temperature of the interface drops down to ~0.92 $T_m$ by 300 ps, and stays close to this level until the end of the solidification process.

The computational prediction that the material redistribution in the lower part of the ablation plume is largely responsible for the generation of the frozen surface protrusion is in conflict with the mechanism of the LIPSS formation suggested in Ref. [17], where a combination of ablative material removal and redistribution of the molten material driven by the gradient of the ablation recoil pressure is suggested to be responsible for the formation of the periodic surface relief. The latter mechanism, *i.e.*, the liquid flow along the surface of the target induced by the pressure gradient, is indeed observed in the simulation, as evidenced by small arrows in the molten (yellow/orange) part of the target in the snapshots shown for 25 – 100 ps in Figure 2. The contribution of this mechanism of material redistribution to formation of the final frozen protrusion, however, is relatively small. To quantify this contribution, we have calculated the total amount of liquid passed towards the central part of the system through vertical planes located at positions of 80 and 180 nm along the *x* axis and at any depth larger than 40 nm, *i.e.*, z ≤ -40 nm in Figure 2. This depth of 40 nm roughly corresponds to the ablation depth in the region of the maximum energy deposition. The total amount of the material redistributed through the liquid flow generated within the molten surface layer of the target is found to only account for about 10% of the material in the protrusion (*z* > -35 nm) after complete solidification of the target. Decreasing the cutoff depth in the analysis of the liquid flow to 35 nm (z ≤ -35 nm) yields a larger contribution of 17% to the protrusion. However, this increase is largely attributed to the displacement of liquid droplets partially attached to the surface and moving with a considerable speed towards the center of the computational system.

As noted above, the predicted shape of the frozen protrusion is similar to the shapes of LIPSS generated on Au surface by single femtosecond pulse irradiation [16,31], but is clearly distinct from the highly regular LIPSS produced by scanning Si surface by high repetition rate femtosecond laser pulses [17], where broad hills separated by narrow troughs are generated. In addition to the differences stemming from material-specific response to the laser excitation, the larger spatial modulation of the absorbed fluence used in the experimental study may cause generation of a thicker liquid wall that could exhibit slower decomposition and more substantial flattening of the protrusion prior to the solidification. The partial overlap of the laser pulses, leading to the effective irradiation by several shots per spot may also smoothen the surface topology via re-melting the walls and relocation of the material driven by surface tension. Finally, the nonlinear nature of the laser energy absorption, particularly a drop of reflectivity at high electron temperatures, could also play a role by further sharpening the contrast between regions of higher and lower energy density deposition. As a result, instead of sinusoidal modulation assumed in the simulation, the deposited laser energy may exhibit narrow spikes that could produce deep troughs in the resolidified surface.

It is notable that the patterning associated with the material redistribution toward the regions of the minimum laser energy deposition is in contrast with the results of simulations and experiments where the protrusions induced by interference of two laser beams at lower fluences appear in the regions of the maximum laser energy deposition [32]. The interference technique implies much higher modulation amplitude of laser intensity than in the present study, up to 50%, along the material surface. Although the electronic heat conduction should somewhat smooth the temperature profile on the temporal scale of the protrusion formation, in the irradiation regime investigated in Ref. [32] the extended regions between the absorption maxima remain solid, thus confining the molten material. As a result, the lower-energy patterning is produced through melting and expansion of regions of the maximum energy deposition, where the generation of subsurface voids captured by rapid solidification [33] is responsible for surface swelling or formation of frozen surface features. In contrast, the material redistribution in the ablative surface patterning is driven by the pressure gradients in the initial ablation plume and results in a rapid density "inversion" in the ablation plume – from the initial maxima above the higher energy deposition regions of the target to the material concentration and formation of the liquid wall in the regions above the minima of the energy deposition.

While the density inversion happens within a few tens of picoseconds in the immediate vicinity of the surface, and may be difficult to see from Figure 2, it is more apparent in the higher part of the ablation plume where the density inversion from the regions above the maxima of the energy deposition to the ones above the minima is observed at a longer timescale of 400-500 ps. The material redistribution reflecting in the modulation of density in the ablation plume is illustrated in Figure 4, where the plume density is shown for a region extending from 1 to 3 μm above the irradiated surface. Initially the plume density is higher in the regions located above the maxima of the laser energy deposition, while later the lateral vapor pressure gradient drives the material redistribution to the central part of the system, leading to the density inversion occurring between 400 and 500 ps. The maximum observed density contrast at a height of 1-1.5 μm above the target is approximately half of the maximum value that reaches 0.5% of the original density of solid Cr target. As time progresses, the plume becomes uniform and, by the time of 1 ns, any lateral density contrast disappears.

Turning the attention to the internal structure of the frozen surface protrusion and the surrounding subsurface region, we note a high density of crystal defects generated in the course of laser-induced melting and resolidification. The distribution of crystal defects is visualized in three complementary images shown in Figure 5. The dislocation configurations can be seen in Figure 5a, where the atoms with elevated potential energy are shown, and the chains of such atoms correspond to dislocation cores. The dislocation types are identified with the algorithm developed in Ref. [34] and are shown In Figure 5b. Analysis of the dynamics of the dislocation generation reveals that they are not produced during the initial dynamic relaxation of laser induced stresses, when the stresses partially relax due to the transient appearance of a high density of stacking faults along the {110} planes [21]. The stacking faults are unstable and quickly disappear when the tensile stress wave leaves the surface region. The emission of the dislocations is then initiated at approximately 500 ps, when the solidification front reaches the base of the liquid wall, and the resolidified material gets an opportunity to expand laterally and generate shear stresses sufficiently strong for the emission of dislocations. The majority of dislocations have a Burgers vector $1/2\langle 111\rangle$, although several segments of $\langle 100\rangle$ dislocations, produced by dislocation reactions, are also observed. The average density of dislocations in the subsurface region of the target included in the MD part of the computational cell is approximately $10^{15}$ m$^{-2}$.

In addition to the dislocations, a very high concentration of vacancies and vacancy clusters is observed inside the frozen protrusion and in a subsurface region of the target, as shown in Figure 5c. The concentrations of vacancies and divacancies in the part of the target that experienced the laser-induced melting and rapid resolidification are calculated to be approximately 0.21% and 0.017% of lattice sites, respectively. The high vacancy concentration is consistent with values of the vacancy concentration observed in earlier simulations of laser processing of Cr targets [21], and largely exceeds the equilibrium vacancy concentration at the melting point, which can be estimated to be approximately $10^{-4}$ of lattice sites. The rapid solidification proceeding under conditions of strong undercooling results in the generation of strong vacancy supersaturation, while the rapid (up to ~$10^{12}$ K/s [21,33]) cooling of the resolidified target does not leave time for equilibration of the vacancy concentration. The estimated vacancy diffusion path is on the order of several nanometers during the rapid quenching of the resolidified surface region, and even the vacancies located close to the free surface can be "frozen" in the target. In addition to the vacancies generated in the course of the rapid solidification, the dislocation reactions and non-conservative movement of dislocations leads to the generation of additional vacancies in the deeper part of the target, below the maximum melting depth (Figure 5c). The average concentration of vacancies generated in this region due to motion of dislocations is approximately 0.012% of lattice sites.

## 4. Summary

A large-scale molecular dynamics simulation is employed to study the mechanisms of single-pulse LIPSS formation in the regimes of strong ablation. The simulation of spatially modulated ablation induced by irradiation at an average laser fluence well above the threshold for the phase explosion reveals a complex interplay of material removal and redistribution leading to the formation of prominent surface features extending above the level of original surface of the target. The fast heating of Cr target by the laser pulse leads to an explosive decomposition of about 30-40 nm deep surface layer into a mixture of vapor and liquid droplets. Lateral pressure gradients in the plume generated by the spatially modulated laser ablation drive the vapor and liquid droplets to the region located above the minima of the laser energy deposition at the target surface. The material redistribution leads to formation of a high-density region evolving into an elongated liquid wall extending up to ~600 nm above the surface of the target. The upper part of the liquid wall disintegrates into droplets while the base of the wall solidifies on the timescale of ~2 ns, producing a ~100 nm-long frozen surface feature.

In contrast to earlier simulations of low-fluence laser processing through spatially modulated energy deposition by two interfering laser beams, where the frozen surface protrusions are generated in regions of the maximum energy deposition, the ablative laser patterning investigated in this work reveals a density "inversion" in the ablation plume, where the maximum plume density rapidly switches from the regions above the maxima to the ones of minima of the energy deposition due to an active lateral motion of the material. The results of the simulation also suggest that the material redistribution in the lower part of the ablation plume is playing the major role in the formation of the frozen protrusions while the liquid flow in the molten part of the target driven by the gradient of the recoil pressure makes a relatively small contribution to the LIPSS formation.

The properties of the solid protrusions and adjacent regions of the target can be expected to be strongly affected by a high vacancy concentration produced through the rapid solidification under conditions of deep undercooling and predicted to exceed $10^{-3}$ of lattice sites. Moreover, the simulation demonstrates generation of a high density of dislocations, $10^{15}$ m$^{-2}$, that can further modify the mechanical and chemical properties of the laser-processed surface. The insights into the LIPSS formation mechanisms may help in designing approaches for increasing the processing speed and improving the quality of the laser-generated surface structures.

**Acknowledgment**

Financial support for this work was provided by the National Science Foundation (NSF) through Grant DMR-1610936. N.M.B. acknowledges the European Regional Development Fund and the state budget of the Czech Republic (project HiLASE CoE: Grant No. CZ.02.1.01/0.0/0.0/15_006/0000674), the European Union's Horizon 2020 research and innovation programme under grant agreement No. 739573, and support from the Ministry of Education, Youth and Sports of the Czech Republic (Programmes NPU I Project No. LO1602, and Large Research Infrastructure Project No. LM2015086). Computational support was provided by the Oak Ridge Leadership Computing Facility (INCITE project MAT130) and NSF through the Extreme Science and Engineering Discovery Environment (project TG-DMR110090).

**Appendix A: Langevin nonreflecting boundary condition**

In MD simulations of laser interaction with a bulk target, the region of the target that can be represented with atomic resolution is typically small and does not exceed several hundreds of nanometers. The propagation of laser-generated stress waves from the area of the laser energy deposition to the bulk of the target can be effectively mimicked with utilization of so called Non-Reflecting Boundary (NRB) conditions [35,36]. This approach, however, is only applicable for simulation of non-reflecting propagation of planar [20,21,27] or spherical/circular [37,38] waves. The pressure wave generated by the spatially modulated laser energy deposition considered in the present work cannot be expected to have a planar front, which makes the traditional NRB inapplicable. To model the nonreflecting propagation of the nonplanar pressure wave, we designed an alternative formulation of NRB based on the Langevin equation, which is briefly descried below.

The basic idea of NRB is to ensure the absorption of incoming elastic waves through the application of an external dumping force to boundary atoms, $F_{NRB} = -ZS_a v$, where $Z$ is the acoustic impedance of the target or substrate material, $S_a$ is the effective area per each boundary atom (the area of the boundary divided by the total number of boundary atoms), and $v$ is the velocity of atoms perpendicular to the boundary. In the traditional NRB [20,21,35] all atoms move together, and the atomic velocities and forces are averaged over all boundary atoms. In contrast, to model nonreflecting propagation of a wave with a nonplanar front, we can treat motion of each boundary atom independently [36]. One difficulty with this approach, however, is that the application of the dumping force to each atom independently not only affects the collective atomic motion associated with the stress wave but also effectively dampens the thermal motion of the boundary atoms. Due to the strong cooling of the boundary region, this approach is only applicable under conditions where the temperature near the boundary is zero [36]. In order to minimize the computational cost of the simulations, however, it is preferable to limit the MD part of the model to a region where the laser-induced structural and phase transformations take place, while the heating and heat transfer that do not cause structural changes can be described at the continuum level. This requirement means that the temperature increase can be substantial at the location of the nonreflecting boundary. As a result, the velocity $v$ corresponds not only to a coherent motion of the boundary layer but also to a thermal motion of individual atoms, which should not be damped. In order to compensate for the loss of the thermal energy, the equation of motion for a boundary atom $i$ can be expressed in a form similar to Langevin dynamics equation with the dumping constant $ZS_a$:

$$m\frac{d^2 z_i}{dt^2} = F_i^z - ZS_a v_i^z + \eta(t),$$

$$\langle \eta(t)\eta(t') \rangle = 2ZS_a k_B T \delta(t-t'),$$

where the boundary is assumed to be perpendicular to $z$ direction, $m$ is the atomic mass, $v_i^z$ and $F_i^z$ are $z$ components of the velocity and force of atom $i$, with the force originating only from interatomic interactions between the boundary atom $i$ and all the atoms located further away from the boundary than the atom $i$, $\eta$ is the a delta-correlated stationary Gaussian process with zero mean and variance defined by the second equation, and $T$ is the average temperature of the boundary atoms.

Since a high value of dumping force may cause instability of the boundary, this value can be reduced by increasing the number of monoatomic layers included in the Langevin Non-Reflecting Boundary (LNRB) layer, which reduces the effective atomic area $S_a$. In the simulation reported in this paper, the LNRB thickness of three monoatomic layers is found to ensure stable behavior of the boundary condition.

The results of testing of the performance of the LNRB are shown in Figure A1. The first test simulation is performed for a Cr system with dimensions of 4.1 nm × 4.1 nm × 19 nm, periodic boundary conditions applied in lateral directions, free boundary used at the top, and the LNRB imposed at the bottom of the computational system. A strong plane compressive wave with magnitude of 19 GPa is generated at one end of a 19-nm-deep Cr slab by applying an 800 m/s extra velocity to a 1.5-nm-thick top surface layer. As can be seen from the contour plot shown in Figure A1a, the compressive wave propagates with relatively weak reflection, where the reflected wave remains within the range of ±1 GPa. The reflection can be further minimized by introducing pressure dependence of the acoustic impedance $Z$. The performance of the LNRB for nonplanar waves is illustrated in (b), where a sequence of pressure plots the large scale atomistic TTM-MD modeling with a periodically modulated absorption pattern, discussed in the present paper, is shown. Overall, we can conclude that the modified formulation of the boundary condition provides an efficient and simple way to simulate nonreflecting propagation of nonplanar waves.

**Figures and figure legends**

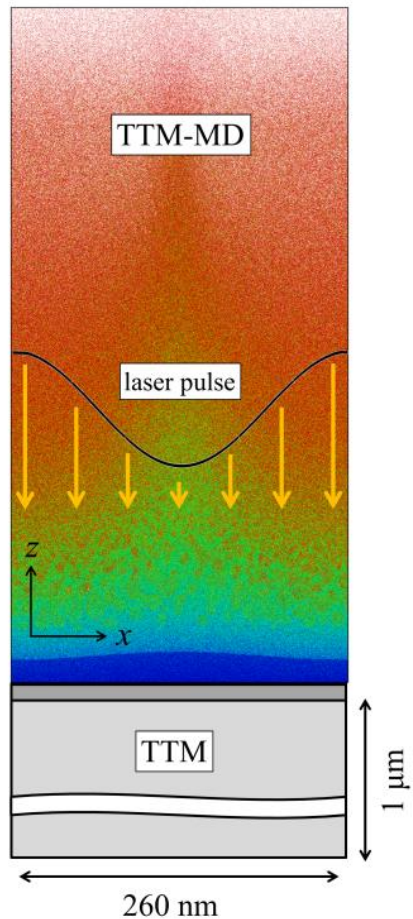

**Figure 1.** Schematic representation of the computational setup used in the simulation of single pulse ablative LIPSS formation. A snapshot from a simulation taken at 80 ps after the laser pulse is used as background in the representation of the atomistic (TTM-MD) part of the model. The atoms in the snapshot are colored by their potential energies with color scale ranging from -3.8 eV (blue) to -1.4 eV (red). The black curve shows schematically the modulation of the laser energy deposition along the $x$ direction, with the minimum at the center of the computational domain.

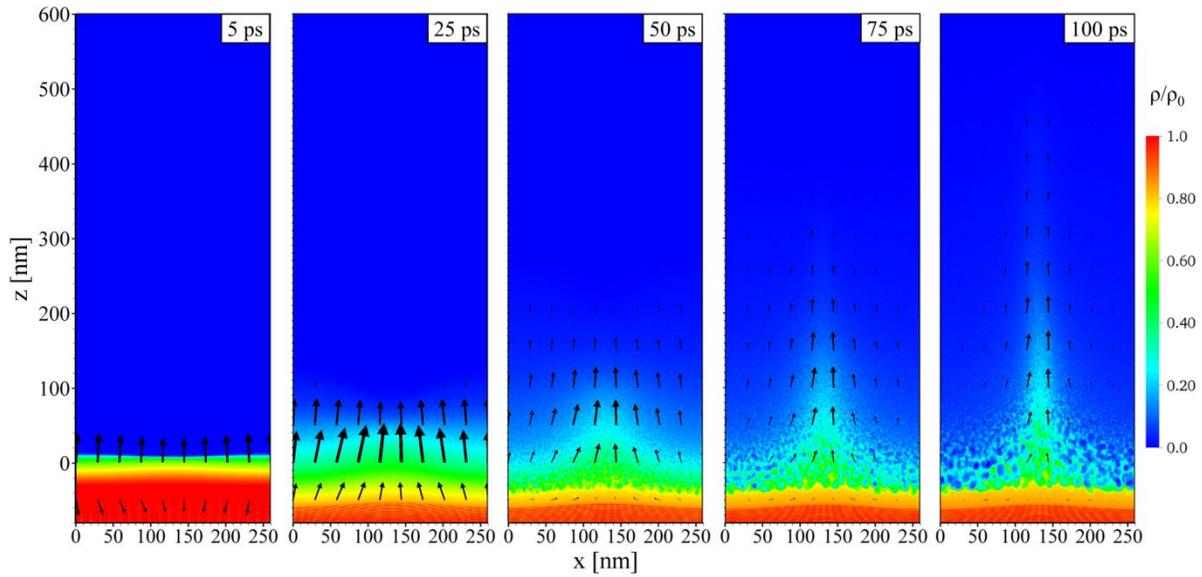

**Figure 2.** The dynamics of material redistribution at the initial stage of the ablation process. Plots are colored based on the instantaneous density distribution, and arrows depict the relative magnitude of material flow (length/thickness of the arrows are scaled by the product of local density and velocity). Red color corresponds to solid, while orange, yellow, and green depict molten material.

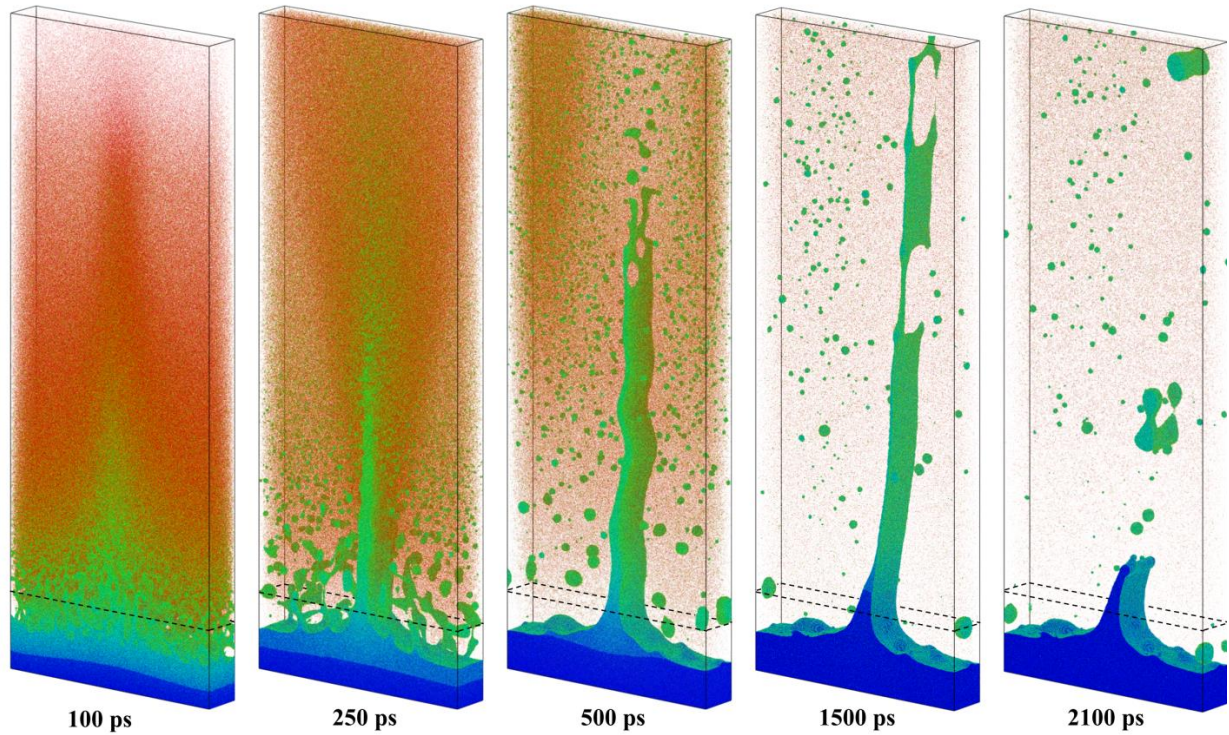

**Figure 3.** A sequence of atomistic snapshots demonstrating formation of a liquid wall in the ablation process followed by rapture of the wall and generation of a solidified protrusion. The snapshots show a part of the system with *z* coordinate ranging from -80 nm to 600 nm with respect to the location of the original surface marked by the dashed lines. The atoms in the snapshots are colored by their potential energy, with color scale ranging from -3.8 eV (blue) to -1.4 eV (red). With this coloring scheme, dark-blue regions correspond to the solid phase, light-blue and green represent liquid phase and free surfaces, and red atoms belong to the vapor phase. The dashed line outlines the initial position of the surface.

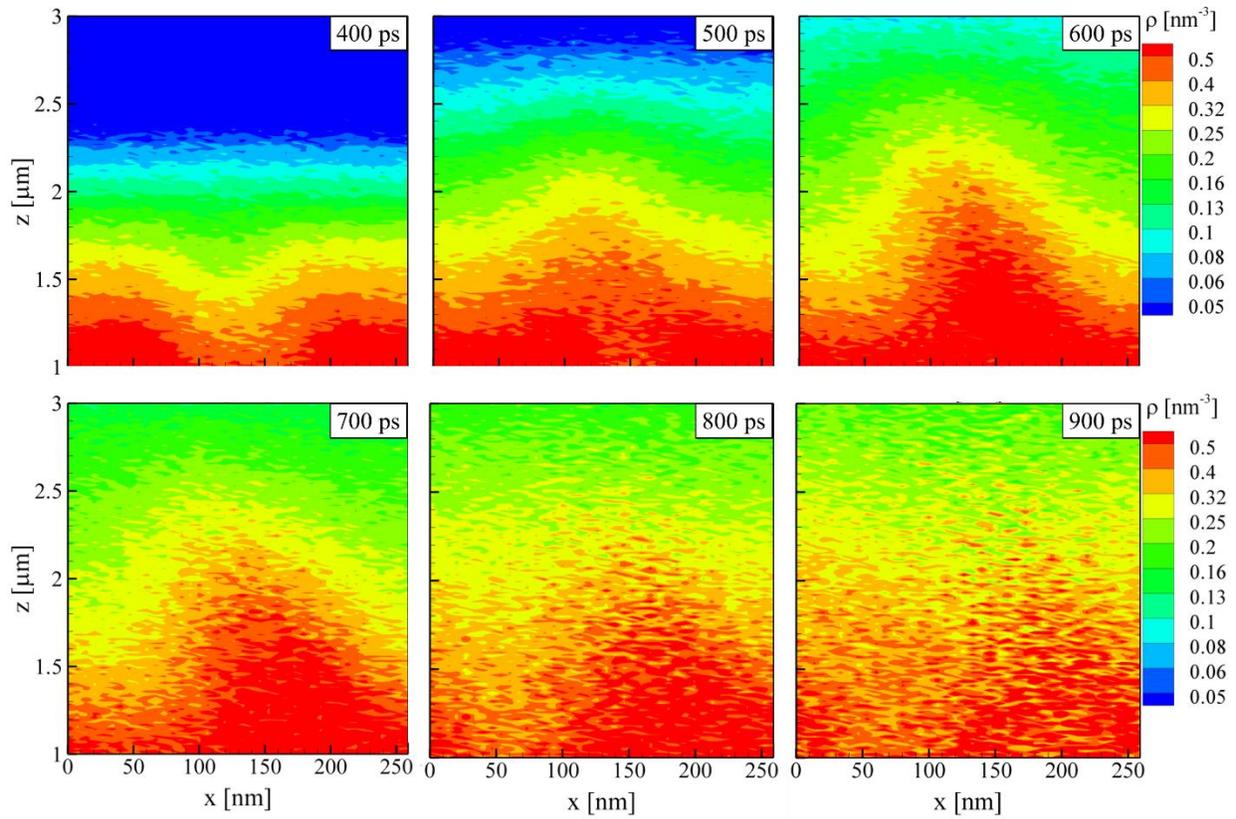

**Figure 4**. Density evolution in the ablation plume generated in a simulation of spatially modulated ablation illustrated by Figures 2 and 3. The plume density is expressed in units of number of Cr atoms per nm$^3$ and is shown for a region extending from 1 to 3 μm above the irradiated surface.

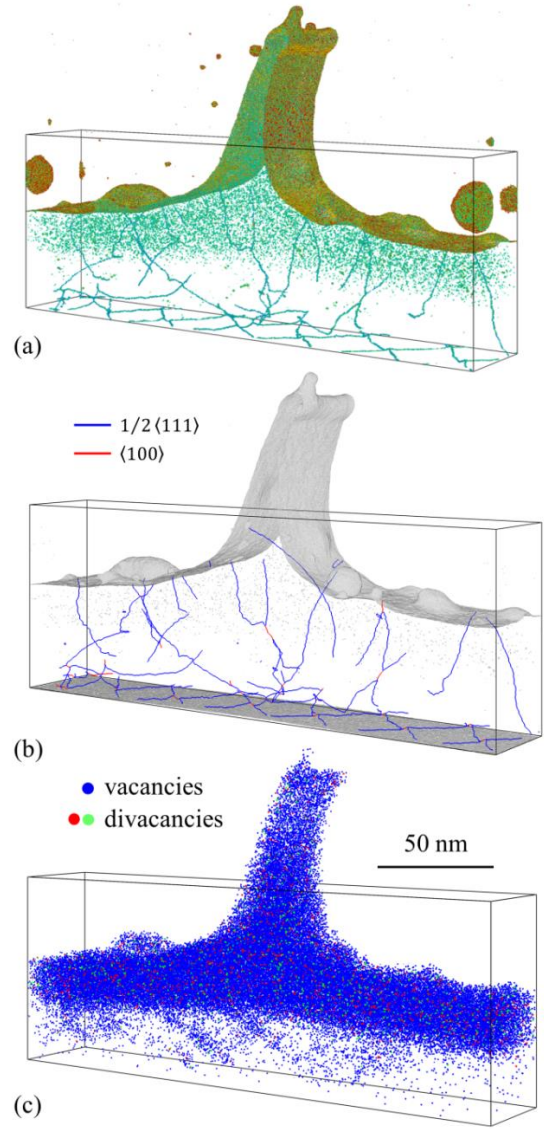

**Figure 5.** Three representations of the microstructure of the surface region of the irradiated Cr target at 2100 ps, after complete resolidification of the target. In (a), dislocations and vacancy clusters present in and below the frozen protrusion are exposed by blanking the atoms with the potential energy below -3.94 eV (correspond to atoms with BCC local coordination and atoms surrounding single vacancies) and the remaining atoms are colored by their potential energy, with color scale ranging from -4.0 eV (blue) to -3.4 eV (red). In (b), dislocations identified with the algorithm of Ref. [34] are shown and colored by dislocation type. In (c), the distribution of vacancies and vacancy clusters is shown, with single vacancies marked by blue color while divacancies with two vacancies on the first and second nearest neighbor lattice sites with respect to each other marked green and red, respectively. The box outlines the shape of the original target before the laser irradiation. Before analysis, the system is quenched for 1 ps using the velocity-dampening technique.

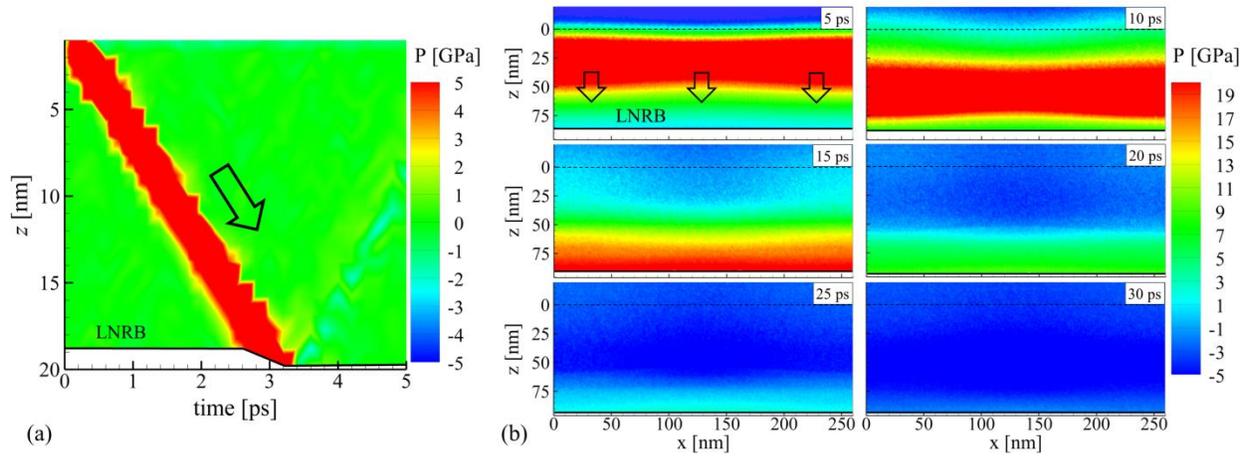

**Figure A1**. The results of testing the LNRB performance in a simulation of a nonreflecting propagation of an 18-GPa-strong compressive planar wave (a) and in a TTM-MD simulation of laser ablation described in this paper, where a nonplanar wave is generated (b). In (a), a free boundary condition is applied at the top of the system, at $z = 0$, and the LNRB is initially located at $z = 19$ nm. In (b), the horizontal dashed line shows the initial position of the surface of the irradiated Cr target, and the LNRB is located at $z = 87$ nm. In both panels, the arrows show the direction of the pressure wave propagation.